\documentclass[aps,prl,showpacs,twocolumn]{revtex4}
\usepackage{graphicx}
\usepackage{amsmath}
\usepackage{color}

\def\bp{{\bf p}}

\input{epsf}
\begin{document}

\title{Radiation-induced quantum interference in low-dimensional $n$-$p$ junctions}

\author{M. V. Fistul$^1$, S. V. Syzranov$^1$, A. M. Kadigrobov$^{1,2}$, and K. B. Efetov$^{1}$}
\affiliation{$^1$Theoretische Physik III, Ruhr-Universit\"at
Bochum, D-44801 Bochum, Germany\\
$^2$ Department of Physics, University of Gothenburg, SE-412 96
G{\" o}teborg, Sweden }
\date{\today}

\begin{abstract}
We predict and analyze {\it radiation-induced quantum interference
effect} in low-dimensional $n$-$p$ junctions. This phenomenon
manifests itself by large oscillations of the photocurrent as a
function of the gate voltage or the frequency of the radiation.
The oscillations result from the quantum interference between two
electron paths accompanied by resonant absorption of photons. They
resemble Ramsey quantum beating and Stueckelberg oscillations
well-known in atomic physics. The effect can be observed in one-
and two-dimensional $n$-$p$ junctions based on nanowires, carbon
nanotubes, monolayer or bilayer graphene nanoribbons.
\end{abstract}

\maketitle


Although quantum mechanics was born more than a century ago, only
in the last decades it became possible to manipulate in a coherent
way the states of single discrete-level systems.
Such fundamental phenomena as microwave-induced Rabi oscillations
\cite{Rabi} and Ramsey quantum beating \cite{Ramsey}, which are
well known in atomic physics and in principle can be realized in
any two-level system, were observed
recently in solid-state devices, for example, in Josephson qubits
\cite{VionMartinis}, quantum dots \cite{Kouwenhoven}, and
ferromagnetic domain walls \cite{Nowack}.
In order to manipulate quickly the states of a macroscopic system
one normally applies a time-dependent perturbation, e.g.
irradiates the system by an electromagnetic field (EF), with the
frequency close to the splitting between respective energy levels.
Time-dependent coherent phenomena in solid-state systems are
observed traditionally in zero-dimensional systems, such as
quantum dots or qubits.

Remarkably, the states of moving electrons in a two-band
semiconductor can be handled by means of external radiation
analogously to those of a two-level system. Indeed, if an electron
propagates in a semiconductor in presence of a non-uniform
potential, then its momentum $\bp({\bf r})$ is coordinate-dependent,
and so is the splitting $\epsilon_c(\bp)-\epsilon_v(\bp)$ between
the energies in the conduction and the valence bands.
 The interaction of
the electron with the radiation occurs effectively only in the
``resonant regions'', near the points where the resonant condition
\begin{eqnarray}\label{Rescondition}
    \epsilon_c(\bp_{res})-\epsilon_v(\bp_{res})=\hbar\omega
\end{eqnarray}
is satisfied, $\hbar\omega$ being the photon energy. The
transmission of an electron through a narrow resonant region is
equivalent in its reference frame to the application of a short
resonant pulse, which can coherently transfer the electron from
one band to another. Electron states in the conduction and the
valence band play here the same role as the two states of a
two-level system (a qubit) subject to an EF. The
coordinate-dependent potential and the distribution function of
the incident electrons determine the effective time of the
resonant interaction with the EF and the initial state of the
qubit respectively.

For a particular example of graphene $n$-$p$ junctions, the electron
motion can be considered as the dynamical Landau-Zener tunneling
through the dynamical gap $\Delta_R$ opened in the electron spectrum
by the resonant interaction with the EF~\cite{FEprl,SFEprb}.
 Varying the frequency $\omega$
and the intensity $S$ of applied EF one can suppress transport in
the junction \cite{FEprl,SFEprb} or generate photocurrent, the
directed current flowing without any dc bias applied \cite{SFEprb}.
The opening of the dynamical gap and some of its effects on the
classical bulky properties of semiconductors have been studied since
quite a while ago \cite{GGE}.

Although the dynamical gap $\Delta_R$ bears a remarkable resemblance
to the Rabi frequency \cite{Rabi,Hanggi}, observed routinely in
experiments with two-level systems
\cite{VionMartinis,Kouwenhoven,Nowack},
 there is no obvious
way to vary the time of resonant interaction of electrons with EF,
and the coherent quantum-mechanical transport phenomena in
low-dimensional $n$-$p$ junctions have not been studied yet.
However, the transmission of electrons through the resonant regions
is equivalent to the dynamics of a two-level system subject to a
sequence of resonant pulses. Thus, one may anticipate certain
manifestations of Ramsey oscillations \cite{Ramsey} in the transport
properties of semiconducting junctions.

\begin{figure}[tbp]
\includegraphics[width=\columnwidth,angle=0]{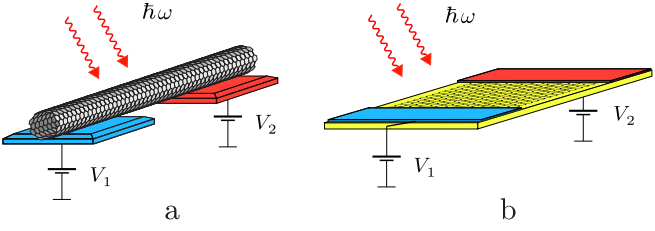}
\caption{\label{setup} (Color online) Schematics of $n$-$p$
junctions irradiated by EF. (a) Carbon nanotube. (b) Monolayer or
bilayer graphene nanoribbon. The applied gate voltage difference
$V_G=V_2-V_1$ allows one to tune the potential profile in the
junction.} \label{Schematic}
\end{figure}
\begin{figure}[tbp]
\includegraphics[width=2in,angle=0]{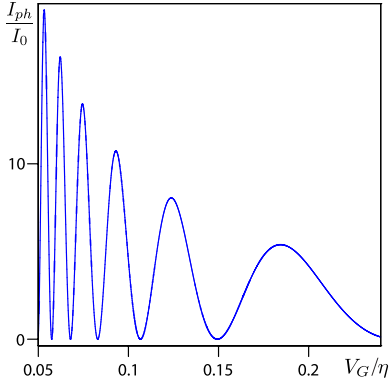} %
\caption{\label{plot} (Color online) The photocurrent $I_{ph}$
displaying oscillations as a function of the gate voltage
difference $V_G$. Here we set $\hbar \omega=1.5E_g$. The notations
$\eta$ and $I_0$ are explained in the text [cf.
Eq.~(\ref{Photocurrent-GNR})].} \label{Fig3}
\end{figure}

In this Letter we predict and analyze the effect of \emph{the
radiation-induced quantum interference} on the ballistic transport
in low-dimensional $n$-$p$ junctions. We show that the quantum
interference leads to large oscillations of the photocurrent
$I_{ph}$ as a function of the difference of the gate voltages
$V_G=V_2-V_1$ (cf. Fig.~\ref{setup}). The dependence $I_{ph}(V_G)$
is shown in Fig.~\ref{plot} and has a universal form; the
parameters of the system and the radiation determine only the
amplitude and period of oscillations. The effect is rather strong
in 1D semiconducting systems, such as nanowires, carbon nanotubes
(CNTs) (Fig.~\ref{setup}a), or graphene nanoribons (GNRs)
(Fig.~\ref{setup}b), but can also be observed in some 2D system,
e.g. based in monolayer or bilayer graphene.

\begin{figure}[tbp]
\includegraphics[width=0.8\columnwidth,angle=0]{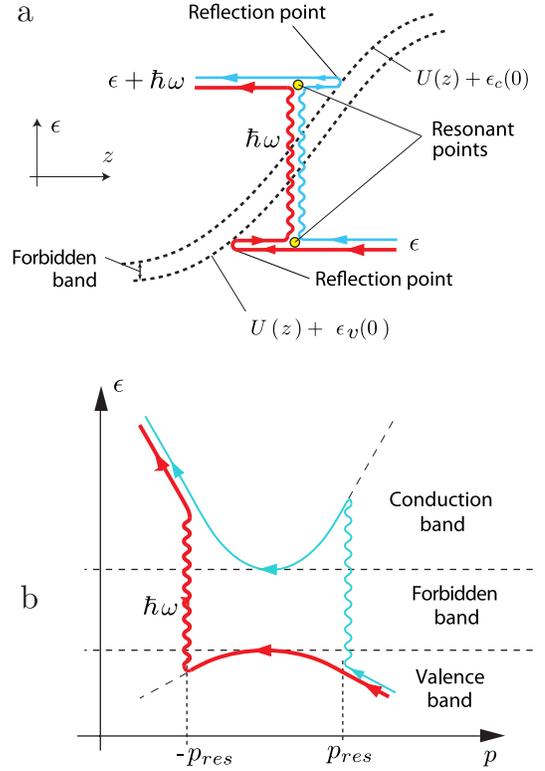} %
\caption{\label{trajectories} (Color online) Typical electron
trajectories in irradiated 1D $n$-$p$ junction. (a) Electron energy
as a function of the spatial coordinate. (b) Energy versus momentum.
The regime $V_g \gg E_g$ is shown.} \label{PhaseTr}
\end{figure}

 The oscillations of the photocurrent
can be understood qualitatively as follows. The transmission of
electrons through the junction is determined by two processes,
namely, by the resonant absorption of photons near the ``resonant
points'', where the condition (\ref{Rescondition}) is satisfied,
and by the strong reflection from the junction interface,
occurring at the ``reflection points'', where the longitudinal
component of momentum  $p_z$ ($z$ is the direction perpendicular
to the interface) turns to zero. The resulting electron
trajectories, which contribute to the photocurrent, are shown in
Fig.~\ref{trajectories}. As one can see, there are two paths
corresponding to the propagation from the right to the left: on
the first one electrons move in the valence band between the
resonant and the reflection points [red (black) line in
Fig.~\ref{trajectories}], while on the second one analogous motion
occurs in the conduction band [blue (gray) line in
Fig.~\ref{trajectories}]. The interference between these two paths
results in the oscillating dependence of the photocurrent on the
gate voltages or on the frequency $\omega$ of the EF.

Let us present the quantitative analysis of the radiation-induced
interference.
 The Hamiltonian of a 1D two-band semiconductor
in presence of external EF and the coordinate-dependent potential
$U(z)$ reads
\begin{eqnarray}
    \hat{H}=[{\epsilon_c(p_z)+\epsilon_v(p_z)}]/{2}+\hat{\sigma}_z[{\epsilon_c(p_z)-\epsilon_v(p_z)}]/{2}
    \nonumber \\
    + 2\Delta_R(p_z)\hat{\sigma}_x \cos(\omega t)+U(z).\label{Hamiltonian}
\end{eqnarray}
Here $\hat{\sigma}_{x,z}$ are the Pauli matrices, and $\Delta_R$
is the matrix element of the resonant interband transitions, that
depends on the intensity $S$ of the EF
($\Delta_R\propto\sqrt{S}$), on its polarization, and on the type
of the semiconducting material. The potential profile $U(z)$ of
the junction can be tuned by applying the gate voltages,
Fig.~\ref{setup}. For simplicity we disregard electron spins and
assume that there is only one valley.


It is convenient to carry out the calculations in the basis of
electron eigenstates, which are related to the initial ones by the
unitary transformation $\hat{V}(t)=\exp(-i\omega t
{\hat{\sigma}_z}/{2})$. The respective transformed Hamiltonian
$\hat{H}^{\prime }=\hat{V}^{+} H\hat{V}- i\hbar \hat{V}^{+}
\hat{\dot{V}}$
 contains static parts and those proportional to
$\exp(\pm 2i\omega t)$. Similarly to the generic case of a
two-level system, one can use the rotating-wave approximation
(RWA) \cite{Hanggi}, i.e. neglect the latter parts of the
Hamiltonian $\hat{H}^{\prime}$ near the resonance. Thus, we obtain
the effective Hamiltonian
\begin{eqnarray}
\hat{H}_{eff}=\left[{\epsilon_c(p_z) +\epsilon_v(p_z)}\right]/{2}
\nonumber \\
+\hat{\sigma}_z\left[{\epsilon_c(p_z)-\epsilon_v(p_z)-\hbar\omega}\right]/{2}
+\Delta_R\hat{\sigma}_x+U(z). \label{Hamiltonian-EFF}
\end{eqnarray}
The same procedure has been used to analyze the electron transport
in a graphene monolayer \cite{FEprl,SFEprb} and spin-dependent
transport in a 2D electron gas \cite{FEspin}. The RWA is valid as
long as the amplitude of the EF is sufficiently small, $\Delta_R \ll
\hbar \omega$. In the absence of the coordinate-dependent potential
$U(z)$ the eigenvalues of $\hat{H}_{eff}$ read
\begin{eqnarray}\label{Eigenvalues}
    E_{\pm}(p_z)=[{\epsilon_c(p_z)+\epsilon_v(p_z)}]/{2}
    \nonumber\\
    \pm\left\{{[\epsilon_c(p_z)-\epsilon_v(p_z)-\hbar\omega]^2}/{4}+\Delta_R^2\right\}^\frac{1}{2}.
\end{eqnarray}
The Eq.~(\ref{Eigenvalues}) shows that the resonant interaction of
electrons with EF opens the dynamical gap $\Delta_R$ in the
electron spectrum at the resonant momentum $p_z=p_{res}$
satisfying condition (\ref{Rescondition}).

Let us proceed to the analysis of electron dynamics in presence of
the coordinate-dependent potential $U(z)$. The classical phase
trajectories $p_z(z)$ of the Hamiltonian $\hat{H}_{eff}$ are
determined by the energy conservation law:
\begin{equation}
U(z)+E_{\pm}(p_z)=\tilde\epsilon, \label{energyconserv}
\end{equation}
where $\tilde\epsilon$ is the electron energy in the transformed
basis, related to that in the initial basis as
$\tilde\epsilon=\epsilon\mp\hbar\omega/2$ in the conduction
(valence) bands far from resonant points.

Electron transmission through the resonant regions is determined by
the tunneling through the dynamical gap $\Delta_R$, analogously to
that through the forbidden band of conventional semiconductors
\cite{TunnKadig} or to the transmission of electrons through a
graphene $n$-$p$ junction \cite{Falko, NovGeim}. The position of the
resonant point $z_0$ is determined by the condition
\begin{equation}
    U(z_0)+[{\epsilon_c(p_{res})+\epsilon_v(p_{res})}]/{2}=\tilde\epsilon,
\end{equation}
and Eq.~(\ref{Rescondition}). Approximating the potential $U(z)$
in the vicinity of the the point $z_0$ by a linear function, one
can describe the transmission through the resonant region by the
Landau-Zener tunneling \cite{SFEprb,FEprl}. The tunneling
probability reads
\begin{equation}
    T=\exp\left[-{\pi\Delta_R^2}/{(\hbar \mathrm{v}F)}\right],
\end{equation}
where $F$ is the slope of the potential $U(z)$ close to the resonant
point, and
\begin{equation}
    \mathrm{v}=({1}/{2})d\left[\epsilon_c(p_z)-\epsilon_v(p_{z})\right]/dp_z|_{p_z=p_{res}}.
\end{equation}
The reflection from the resonant region, that occurs with
probability $1-T$, corresponds to the processes of the photon
emission (absorption) in the initial basis of electron states. The
reflection is accompanied by the velocity reversal (cf.
Fig.~\ref{trajectories}a).

Far from the resonant point electrons weakly interact with the EF.
The resulting classical trajectories in a smooth potential $U(z)$
are shown in Fig.~\ref{trajectories}. As we mentioned before,
there are two paths allowing the penetration from the right to the
left. The total probability of the inelastic electron transmission
reads
\begin{equation}
P_{RL}={T(1-T)}
|e^{i\varphi_I}+e^{i\varphi_{II}}|^2/2,\label{Transmission}
\end{equation}
where $\varphi_{I(II)}$ are the quantum-mechanical phases along
the two paths (cf. Fig.~3). The accumulated phase difference
$\phi=\varphi_I-\varphi_{II}$ is
\begin{equation}
    \phi\approx({2}/{\hbar
    F})\int_0^{p_{res}}\left[\epsilon_c(p_z)-\epsilon_v(p_z)\right]dp_z.
    \label{Phasedifference}
\end{equation}
Here we assume for simplicity that the potential slope $F$ is
constant in a sufficiently large region close to the $n$-$p$
interface. In this case the phase $\phi$ is independent of the
energy $\tilde\varepsilon$. Eq.~(\ref{Phasedifference}) is the
leading-order quasiclassical contribution to the phase difference;
using it we neglect certain contributions of order unity that can
result only in the shift of the oscillations of the photocurrent.

The current through the junction is given by the modified Landauer
formula, allowing for the photon emission/absorption
\cite{Moskaletz,SFEprb}:
\begin{equation}
    I=ge\sum_n\int\frac{d\epsilon}{2\pi\hbar}P_{RL}(\epsilon,\epsilon+n\hbar\omega)
    \left[f_L(\epsilon)-f_R(\epsilon+n\hbar\omega)\right].
    \label{Landauer}
\end{equation}
Here $f_L(\epsilon)$ and $f_R(\epsilon)$ are the distribution
functions in the left and the right leads,
$P_{RL}(\epsilon,\epsilon+n\hbar\omega)$- the probability to
penetrate from the right to the left lead accompanied by the
energy change from $\epsilon$ to $\epsilon+n\hbar\omega$, $g$- the
number of the degrees of freedom not affecting the transport, e.g.
spins, valleys, transverse channels. Taking into account that each
electron in the energy interval $\hbar\omega$ in the valence band
can absorb a photon and penetrate into the conduction band, we
obtain the photocurrent as
\begin{eqnarray}
    I_{ph}={(1/\pi)g}e\omega T(1-T)\cos^2 \phi.
    \label{Photocurrent-gen}
\end{eqnarray}

Since the phase difference $\phi$ depends on the electric field
$F/e=V_G/d$, where $d$ is the length of an $n$-$p$ junction, the
photocurrent $I_{ph}$ displays oscillations as a function of the
gate voltages, i.e. of $V_G$. The oscillations result from the
quantum interference of electron moving in the confined area
between the resonant and reflection points in the conduction and
the valence bands (see Fig.~3). These oscillations are similar to
the Stueckelberg oscillations \cite{Stueckelberg}, the quantum
interference effect occurring in quantum collisions due to the
superposition of two quantum-mechanical pathways. If considered in
the reference frame of moving electrons, the oscillations of the
photocurrent bear resemblance to the ``Ramsey fringes'', the
quantum beating in the population of a two-level system subject to
a sequence of resonant pulses.
 Indeed,
introducing the period of electron motion in the conduction band as
$\delta t=2\int dz[{d\epsilon_c}/{dp_z}]^{-1}$ and the average
kinetic energy $\hbar \overline{\omega}_0=(2/z_0)\int dz [{\hbar
\omega}/{2}-U(z)]$ of confined electrons, we write the phase
difference
as $\phi=|(\omega-\overline{\omega}_0)|\delta t$, and one can
recognize the well-known parameter determining the period of
``Ramsey fringes'' \cite{Ramsey, VionMartinis}.

Let us apply our generic result, Eq.~(\ref{Photocurrent-gen}), to a
particular case of $n$-$p$ junctions based on GNRs and CNTs. These
quasi-1D objects can be metallic or semiconducting. However, as
their gapless modes do not contribute to the photocurrent, we can
consider at sufficiently low frequencies the effective two-band
spectrum \cite{Grnanotubes,Tubes}
\begin{equation}
\epsilon_{c,v}=\pm\sqrt{v_0^2p_z^2+E_g^2/4}. \label{Spectrum}
\end{equation}
The gap $E_g$ appears due
to the transverse momentum quantization.
Calculating the integral over $p_z$ in Eq. (\ref{Phasedifference})
with the spectrum (\ref{Spectrum}) we arrive at the 
dependence of the photocurrent on the gate voltage $V_G$ (for
$|1-T| \ll 1$):
\begin{eqnarray}
I_{ph}=I_0(\eta/V_G)\cos^2 [f({E_g}/{\hbar
 \omega}){\eta}/{V_G}],
\nonumber \\
f(x)=\sqrt{1-x^2}+x^2\ln[(1+\sqrt{1-x^2})/x]
\nonumber \\
\eta=\frac{\hbar\omega^2d}{2ev_0},~~ I_0=\frac{8e\Delta_R^2
}{\hbar^2
\omega}\left[1-\left(\frac{E_g}{\hbar\omega}\right)^2\right]^\frac{1}{2}
\label{Photocurrent-GNR}
\end{eqnarray}
The parameters $I_0$ and $\eta$ determine respectively the amplitude
and the period of the oscillations. The typical dependence $I_{ph}$
on $V_G$ is shown in Fig.~\ref{plot}.

In order to observe the quantum oscillations in the photocurrent a
few conditions are to be satisfied. The photon energy
$\hbar\omega$ must exceed the width of the forbidden band $E_g$.
At the same time, to maximize the effect, $\hbar\omega$ should be
sufficiently small not to involve in the transport any higher
bands, and the forbidden band $E_g$ has to be large enough,
$E_g\geq \sqrt{\hbar v_0 eV_G/d}$ to ensure a strong reflection
from the junction interface. The gate voltage difference $V_G$
should be sufficient, $eV_G\gg E_g$, to create a smooth potential
profile with a nearly constant slope close to the interface. The
order of magnitude of the energy gap $E_g$ can vary from $0.01eV$
in strained metallic carbon nanotubes to $1eV$ in semiconducting
CNTs or in GNRs \cite{Tubes}, so the desired radiation frequency
may be in the THz or in the infrared optical region. For example,
for the typical junction parameters $E_g=0.1eV$, $d=100nm$, and
for the radiation frequency $\nu=50THz$, the characteristic scale
$\eta$ (cf. Fig.~\ref{plot}) estimates $2.5V$. Thus, one can
observe a large number of oscillations at $V_G\lesssim 0.3V$.

Since the oscillations of the photocurrent $I_{ph}$ is a coherent
quantum interference effect, they can be destroyed by disorder and
inelastic processes. For a good visibility of the oscillations the
characteristic time of electron propagation between the resonant and
the reflection points should be smaller than the scattering time,
i.e.  $d/v \ll \tau$. On the other hand the deviation of the
dependency $I_{ph}(V_G)$ from the universal form
[Eq.~(\ref{Photocurrent-GNR})] allows one to quantitatively analyze
the processes of elastic and inelastic scattering in such systems.

The oscillations can be observed in some 2D $n$-$p$ junctions as
well, for example, in those based on bilayer graphene. Electrons
there have a quadratic two-band spectrum $\epsilon(p)=\pm p^2/2m$,
which leads to the strong reflection from the $n$-$p$ interface
\cite{NovGeim,Blanter} and thus allows one to confine electrons
between the resonant and reflection points, as necessary for
observing the oscillations. Their characteristic period is
$\eta_{bg}/V_G$ with $\eta_{bg}=\omega d\sqrt{m\hbar \omega}/e$. In
$2D$ systems the phase $\phi$ depends on the transverse momentum
$p_y$, $\phi_{bg}(p_y)=(\eta_{bg}/V_G) [1-p_y^2/(m\hbar
\omega)]^{3/2}$. As a result, the oscillations of the photocurrent
$I_{dc}~\propto \int dp_y P_{RL}(p_y)$ are smeared and suppressed by
the parameter $\sqrt{\hbar\omega/\eta_{bg}}$, as compared to the 1D
case.

In conclusion, we predict and analyze radiation-induced quantum
interference effect in the dc transport properties of
low-dimensional $n$-$p$ junctions subject to an externally applied
EF. This phenomenon manifests itself by large oscillations in the
dependence of the photocurrent on  the gate voltage or on the
frequency of the EF.
The effect can be observed in diverse quasi-1D semiconducting
systems, such as nanowires, carbon nanotubes (CNTs), or graphene
nanoribons (GNRs), as well as in some 2D system, e.g. bilayer
graphene.

We are grateful to L.S.~Levitov for useful discussions and to SFB
491, SFB Transregio 12, and the European Commission (FP7-ICT-2007-C;
project No. 225955 STELE) for the financial support.

\end{document}